\documentclass[preprint,authoryear,12pt]{elsarticle}

\usepackage{graphics}

\usepackage{amssymb}
\usepackage{amsthm}
\usepackage{amsbsy}

\usepackage{graphicx}
\usepackage{natbib}
\usepackage{amsmath}
\usepackage{setspace}
\usepackage{subeqnarray}

\newcommand\Rey{\mbox{\textit{Re}}}  

\newsavebox{\astrutbox}
\sbox{\astrutbox}{\rule[-5pt]{0pt}{20pt}}

\newcommand\etal{\mbox{\textit{et al.}}}

\usepackage{graphicx}
\usepackage{caption}
\usepackage{subcaption}

\journal{Applied Mathematical Modeling}

\begin{document}

\begin{frontmatter}



\title{An axisymmetric steady state vortex ring model}


\author[label1]{Ruo-Qian Wang\corref{cor1}}
\ead{rqwang@mit.edu}



\address[label1]{Department of Civil and Environmental Engineering, Massachusetts Institute of Technology, MA 02139, USA}

\cortext[cor1]{Corresponding author. Tel.:$+$1-617-253-6595.}

\begin{abstract}
Based on the solution of \cite{Atanasiu2004}, a theoretical model for axisymmetric vortex flows is derived in the present study by solving the vorticity transport equation for an inviscid, incompressible fluid in cylindrical coordinates. The model can describe a variety of axisymmetric flows with particular boundary conditions at a moderately high Reynolds number. This paper shows one example: a high Reynolds number laminar vortex ring. The model can represent a family of vortex rings by specifying the modulus function using a Rayleigh distribution function. The characteristics of this vortex ring family are illustrated by numerical methods. For verification, the model results compare well with the recent direct numerical simulations (DNS) in terms of the vorticity distribution and streamline patterns, cross-sectional areas of the vortex core and bubble, and radial vorticity distribution through the vortex center. Most importantly, the asymmetry and elliptical outline of the vorticity profile are well captured.
\end{abstract}


\end{frontmatter}


\section{Introduction}\label{sec:introduction}

The studies of vortex rings can be traced back more than one and half centuries, when William Barton Rogers, founder of MIT, conducted the first systematic vortex ring experiments \citep{rogers1858formation}. Inspired by his and other pioneers' work, a series of theoretical endeavors have been made for the mathematical description of vortex rings. An early example is the famous Hill's vortex, which assumes that vorticity is linearly proportion to radius within a spherical volume, with potential flow outside \citep{Hill1894}. The assumption is relatively simple, yet it is able to generate realistic-looking streamlines, and is arguably the most popular vortex ring model in applied science and engineering, e.g.\ \cite{Lai2013}. From a different starting point, \citet{Fraenkel1972} analyzed the vortex ring by extending the theoretical solution of a vortex filament to allow for a small finite thickness. To bridge these two models, \citet{Norbury1973} treated the Hill's spherical vortex and Fraenkel's thin ring as two asymptotic members of a series of generalized vortex rings. He then numerically determined a range of intermediate rings, now referred to as the Norbury-Fraenkel (NF) vortex ring family. The NF family delivers more accurate streamlines and more precise vortex ring outlines. However, its linear distribution of vorticity is still unrealistic \citep{Danaila2008}. 

Recently, a solution for viscous vortex rings at low Reynolds numbers ($\Rey$) was obtained by \cite{Kaplanski1999}. They derived a generalized solution to the diffusing viscous vortex ring, by directly solving the axisymmetric vorticity-stream function equations without the nonlinear convection terms. Their solution delineates a donut shape outline and a more realistic Gaussian distribution of vorticity in the radial direction. \cite{Kaplanski2009} extended the solution to turbulence by adopting an effective turbulent viscosity. However, recently \cite{Danaila2008} conducted Direct Numerical Simulations (DNS) and reported elliptical cross-sections and radial asymmetry for the vorticity distribution, which both Norbury-Fraenkel and Kaplanski-Rudi models are unable to capture. Realizing the issue, \cite{Kaplanski2012} added two adjustable parameters to allow an elliptical cross-section, but the radial asymmetry of the vorticity distribution is still amiss due to their symmetric Gaussian distribution.

Laboratory experiments and numerical simulations have also shed light on vortex ring dynamics as reviewed by \cite{Lim1995} and \cite{Shariff1992}. In particular, \cite{Gharib1998} performed experiments in which a piston generated a non-buoyant vortex puff. They observed that if the aspect ratio of piston stroke length $L$ to diameter $D$ was less than about four, the generated vorticity could be incorporated into the head vortex, whereas for larger aspect ratios a trailing stem occurs. The phenomenon is now referred to as the ``pinch-off'', and the critical aspect ratio is called the ``formation number''. Because the trailing stem stops supplying vorticity to the head vortex after the pinch-off, the ``saturated'' head vortex ring in the post-formation stage is relatively stable, and should resemble the steady state situation in Hill's vortex. The experimental results can advance the state-of-the-art by enabling a meaningful comparison with idealized theoretical models and numerical simulations or experiments with a saturated head vortex ring. An example can be found in \cite{Danaila2008}, and the present study is also initiated in the same spirit.

The best theoretical model would be the analytical solution to the Navier-Stokes equations, which accurately depicts the dynamics of the flow but is difficult to derive. As a compromise, a theoretical model could be built on the solution to the Euler equations, which ignores the viscous effect but captures the more important non-linear dynamics of the flow. In the present case, we focus on the steady state axisymmetric Euler equation, the target of which is to solve an elliptical second order partial differential equation. This equation is a particular form of the Grad-Shafranov equation \citep{shafranov1958magneto, grad1958proceedings},  which is related to the magnetostatic equilibrium in a perfectly conducting fluid and is well known in magnetohydrodynamics (MHD). Specifying different forms of the involved arbitrary functions, plasma physicists have derived a series of analytical solutions to this equation, e.g. the simplest solution of the Solov'ev equilibrium \citep{solov1967soviet}, the Herrnegger-Maschke solution \citep{Herrnegger1972,maschke1973exact}, and the recent more general solution by \cite{Atanasiu2004} (hereafter referred to as AGLM). It can be shown that the Hill's spherical vortex corresponds to the Solov'ev equilibrium, yet no counterparts of the more advanced solutions have been explored for vortex dynamics. This encourages us to take advantage of them, especially the AGLM, to reach a more accurate model of vortex rings. Beyond that, a higher accuracy model can also meet the industrial demand to improve the description of vortex structures to replace the aged Hill's vortex, e.g. \cite{Lai2013}. In particular, a more realistic vorticity distribution may lay a better foundation to address the particle-vorticity interactions, which motivates us to make the following study.

As mentioned earlier, existing models have improved progressively to describe more detailed features of real vortex rings.  However, an outstanding issue is the asymmetry of the vorticity distribution. The present paper provides an alternative analysis that is capable of incorporating the asymmetry, by deriving a theoretical model for high $\Rey$ laminar vortex rings based on the solution to the axisymmetric inviscid vortex flow. The governing equations are first presented in section \ref{sec:general_solutions} and shown to relate to the Grad-Shafranov equation. The alternative vortex ring solution is shown in section \ref{sec:vortex_ring_solutions}, and its properties are derived in section \ref{sec:properties}. Then, the model is compared to existing simulation results in section \ref{sec:Comparison_numerical_simulations}. Summary and conclusions are given in section \ref{sec:summary_conclusion}.
 
\section{A particular solution}\label{sec:general_solutions}
We adopt the cylindrical coordinate system $(r, \phi, x)$, where $x$ is the local longitudinal coordinate. The incompressible axisymmetric vorticity transport equations can be stated as \citep{Fukumoto2008}
\begin{equation} \label{eq:full}
{\partial \xi \over\partial t}+{\partial\over\partial r}(v\xi)+{\partial\over\partial x}(u\xi)={\mu\over\rho}\left( {\partial^2\xi\over\partial x^2} + {\partial^2\xi\over\partial r^2} + {1\over r}{\partial\xi\over\partial r}-{\xi\over r^2} \right)
\end{equation}
where $t$ is time, $\xi$ is the vorticity, $u$ and $v$ are velocity components in $x$ and $r$ directions respectively, $\mu$ is the dynamic viscosity, and $\rho$ is the density.

Introducing the Stokes stream function $\psi$, the velocity components can then be expressed as
\begin{equation}\label{eq:upsi}
u=-{1\over r}{\partial\psi\over\partial r}+U_x, \quad v={1\over r}{\partial\psi\over\partial x},
\end{equation}
where $U_x$ is the translational velocity of the vortex centroid. The vorticity can be defined by the stream function as
\begin{equation} \label{eq:xpsi}
{\partial^2\psi\over\partial r^2}+{\partial^2\psi\over\partial x^2}-{1\over r}{\partial\psi\over\partial r}=-r\xi,
\end{equation}
which provides the closure to solve (\ref{eq:full}). Equations (\ref{eq:full})-(\ref{eq:xpsi}) can therefore be applied to any axisymmetric flows in theory. Substituting (\ref{eq:upsi}) into (\ref{eq:full}) and normalizing the variables by a length scale $l$ and a velocity scale $U$, i.e.\
\begin{equation}
u=Uu^*, \quad v=Uv^*,\quad r=lr^*,\quad x-U_xt=lx^*,\quad \psi=l^2U\psi^*,\quad\xi=Ul^{-1}\xi^*,
\end{equation}
a non-dimensionalized form of (\ref{eq:full}) can be derived as
 
\begin{multline} \label{eq:nond}
-U_x{\partial \xi^* \over\partial x^*} + {\partial\over\partial r^*}\left({\xi^*\over r^*}{\partial\psi\over\partial x^*}\right)-{\partial\over\partial x^*}\left({\xi^*\over r^*}{\partial\psi\over\partial r^*}\right) + {\partial\over\partial x^*}\left(U_x\xi^*\right)\\
={1\over{\Rey}}\left( {\partial^2\xi^*\over\partial x^{*2}} + {\partial^2\xi^*\over\partial r^{*2}} + {1\over r^*}{\partial\xi^*\over\partial r^*}-{\xi^*\over r^{*2}} \right),
\end{multline}
where $\Rey = \rho Ul/\mu$. $U$ and $l$ can be specified for a particular application. Note that if $\Rey \ll 1$, the nonlinear terms on the LHS of (\ref{eq:nond}) can be neglected, leading to the theoretical solution obtained earlier by \cite{Kaplanski1999}.

For $\Rey \gg 1$ but before transition into turbulence, the viscous effect (RHS of equation (\ref{eq:nond})) can be ignored in the high Re laminar flows. Assuming a steady translation such that $U_x$ is constant, and dropping the * sign for brevity, then
\begin{equation}\label{eq:sxps}
{\partial\over\partial r}\left({\xi\over r}{\partial\psi\over\partial x}\right)-{\partial\over\partial x}\left({\xi\over r}{\partial\psi\over\partial r}\right)=0.
\end{equation}
Using $\xi=rf(\psi)$ \citep{Shariff1992} and substituting into (\ref{eq:xpsi}), we get
\begin{equation} \label{eq:focus}
{\partial^2\psi\over\partial r^2}+{\partial^2\psi\over\partial x^2}-{1\over r}{\partial\psi\over\partial r}=-r^2f\left(\psi\right).
\end{equation}
As mentioned earlier, equation (\ref{eq:focus}) is a particular form of the Grad-Shafranov equation:
\begin{equation} \label{eq:gf}
{\partial^2\psi\over\partial r^2}+{\partial^2\psi\over\partial x^2}-{1\over r}{\partial\psi\over\partial r}=-r^2f\left(\psi\right)+g\left(\psi\right),
\end{equation}
with the arbitrary function $g(\psi)=0$. 

The simplest particular solution to (\ref{eq:gf}) is the Solov'ev equilibrium solution, which assumes
\begin{equation} \label{eq:hill0}
f(\psi)=A, \quad g(\psi)=B
\end{equation}
where $A$ and $B$ are constants. If $B=0$, this reduces to the Hill's spherical vortex solution \citep{Hill1894}, which can be solved in spherical coordinates by separation of variables, while confining the vorticity within the spherical vortex boundary. Although Hill's streamlines are similar to experimental and numerical observations, the vorticity distribution is too simplified and, as will be further discussed, differs significantly from observations \citep{Danaila2008}.

In MHD, more solutions are available and have the potential to improve the accuracy over the simplest Hill's spherical vortex. One example is the particular solution with $f(\psi)=A\psi$ and $g(\psi)=B\psi$, which is known as the Herrnegger-Maschke solution \citep{Herrnegger1972,maschke1973exact}. In addition, a more general solution of AGLM by \cite{Atanasiu2004} covers both situations above, assuming a series expansion of $f(\psi)$ and $g(\psi)$, i.e. 
\begin{equation} \label{eq:Atan}
f(\psi)=A\psi+B,\quad
g(\psi)=C\psi+D, 
\end{equation}
where $A,B,C$ and $D$ are constants.

The AGLM solution inspires us to generate a counterpart model of vortex rings as a meaningful extension to the Hill's spherical vortex. If $C=D=0$, equation (\ref{eq:focus}) becomes
\begin{equation} \label{eq:aglm}
{\partial^2\psi\over\partial r^2}+{\partial^2\psi\over\partial x^2}-{1\over r}{\partial\psi\over\partial r}=-Ar^2\psi-Br^2.
\end{equation}
Thus, the Hill's vortex solution can be viewed as the zeroth order approximation, $A=0$, which linearizes the governing equations for solvability. Note that higher order series expansions can be used in the source term of (\ref{eq:focus}), which involves the nonlinear term and no explicit solution is available so far.

Recalling that $\xi=rf(\psi)$, the improvement of (\ref{eq:Atan}) over (\ref{eq:hill0}) is the manner in which the vorticity $\xi$ correlates with the stream function $\psi$. As discussed earlier, Hill's model delivers realistic-looking stream lines, and the linear vorticity distribution is also advantageous for the handling of the viscous terms in (\ref{eq:nond}) and thus becomes the only explicit solution to the Beltrami flow \citep{Wang1991}.  However, the vorticity distribution is unrealistic. The present solution of (\ref{eq:Atan}) assumes that the vorticity distribution is the first radial moment of the stream function, and by doing so directly correlates the vortex and flow structures. It will be shown later that an improved matching of the stream function and vorticity distribution can be obtained in this manner. Furthermore, the correlation better reflects the physics that the vorticity is transported by the flow structure. 

In the following, we will show how the explicit exact solution can be sought with this first order approximation, which is similar to \cite{Atanasiu2004} but with improvements in details. The inhomogeneous nature of (\ref{eq:aglm}) leads to  a solution consisting of two parts, i.e. $\psi=\psi_o+\psi_{no}$, where $\psi_o$ and $\psi_{no}$ are the homogeneous and inhomogeneous parts, respectively. We shall obtain a general solution for the homogeneous problem first, and then complete the solution with a particular inhomogeneous solution.

Taking the Fourier transform of the homogeneous problem with respect to $x$, we have
\begin{equation}\label{eq:fourier}
{\partial^2\hat{\psi}\over\partial r^2}-\frac{1}{r}\frac{\partial\hat{\psi}}{\partial r}+(Ar^2-\lambda^2)\hat{\psi}=0,
\end{equation}
where the Fourier transform is defined by
\begin{equation}
\hat{\psi}=\int_{-\infty}^{\infty}\psi_o e^{-i\lambda x}dx.
\end{equation}


Equation (\ref{eq:fourier}) can be simplified by introducing $\eta=\alpha r^2$ 
\begin{equation}
\frac{{{\partial ^2}\hat \psi }}{{\partial {\eta ^2}}} + \left( { - \frac{1}{4} - \frac{{{\lambda ^2}}}{{4\alpha \eta }}} \right)\hat \psi  = 0,
\end{equation}
which has a solution of
\begin{equation}
\hat{\psi}=C_{\lambda 1}M_{\frac{-\lambda^2}{4\alpha},\frac{1}{2}}(\alpha r^2)+C_{\lambda2}W_{\frac{-\lambda^2}{4 \alpha},\frac{1}{2}}(\alpha r^2)
\end{equation}
where $M_{k,m}(x)$ and $W_{k,m}(x)$ are the Whittaker $M$ and $W$ functions, $i$ is the imaginary unit, and $\lambda$, $C_{\lambda 1}, C_{\lambda 2}$ are constants to be determined, $\alpha=\sqrt{A}i$ if $A>0$ and $\alpha=\sqrt{-A}$ if $A<0$ \citep{Polyanin2003}. The solution in AGLM is also consistent with the present result but has a much lengthy form. In other words, the present solution is more compact and succinct.

When $A>0$, the solution can also be expressed as:
\begin{equation}
\hat{\psi}={C_{\lambda 1}}{F_0}\left( {\frac{{{\lambda ^2}}}{{4\sqrt A }},\frac{{\sqrt A {r^2}}}{2}} \right) + {C_{\lambda 2}}{G_0}\left( {\frac{{{\lambda ^2}}}{{4\sqrt A }},\frac{{\sqrt A {r^2}}}{2}} \right),
\end{equation}
where  $F_L(\eta,x)$ and $G_L(\eta,x)$ are the Coulomb wave functions of the first and the second kind. This solution is consistent with the Herrnegger-Maschke solution.

%

Subsequently, the stream function and vorticity can be recovered by taking the inverse Fourier transform as
\begin{equation}
\psi_o=\int_{-\infty}^{\infty}\left[C_{\lambda 1}M_{-\frac{\lambda^2}{4\alpha},\frac{1}{2}}(\alpha r^2)+C_{\lambda2}W_{-\frac{\lambda^2}{4\alpha},\frac{1}{2}}(\alpha r^2)\right][\cos(\lambda x)-i\sin(\lambda x)]d\lambda.
\end{equation}

The inhomogeneous solution $\psi_{no}$ should satisfy the equation of
\begin{equation}
{\partial^2\psi_{no}\over\partial r^2}+{\partial^2\psi_{no}\over\partial x^2}-{1\over r}{\partial\psi_{no}\over\partial r}=-Ar^2\psi_{no}-Br^2.
\end{equation}
The simplest particular solution is $\psi_{no}=-B/A$. To summarize, the solution to the reduced Grad-Shafranov equation is
\begin{equation}\label{eq:solp}
\psi=\int_{-\infty}^{\infty}\left[C_{\lambda 1}M_{-\frac{\lambda^2}{4\alpha},\frac{1}{2}}(\alpha r^2)+C_{\lambda2}W_{-\frac{\lambda^2}{4\alpha},\frac{1}{2}}(\alpha r^2)\right][\cos(\lambda x)-i\sin(\lambda x)]d\lambda-B/A,
\end{equation}
and
\begin{equation}\label{eq:solx}
\xi=Ar\int_{-\infty}^{\infty}\left[C_{\lambda 1}M_{-\frac{\lambda^2}{4\alpha},\frac{1}{2}}(\alpha r^2)+C_{\lambda2}W_{-\frac{\lambda^2}{4\alpha},\frac{1}{2}}(\alpha r^2)\right][\cos(\lambda x)-i\sin(\lambda x)]d\lambda.
\end{equation}

In theory, (\ref{eq:solp}) and (\ref{eq:solx}) apply to any axisymmetric inviscid flow in steady state (i.e. with constant translation $U_x$) so long as the first order approximation in (\ref{eq:Atan}) is adequate. We now develop a theoretical model of the high $\Rey$ laminar vortex ring using these solutions together with the necessary boundary conditions.

\section{Vortex ring model}\label{sec:vortex_ring_solutions}
 
We seek an axisymmetric vortex ring solution with a finite boundary that is also symmetric w.r.t. the plane $x=0$, i.e. 
\begin{subeqnarray}
\psi=const. \quad(r=0),\label{eq:BCa}\\
\frac{\partial\psi}{\partial x}=0\quad(x=0),\label{eq:BCb}\\
\psi=const. \quad(at~ \Omega)\label{eq:BCc},
\end{subeqnarray}
where the boundary of $\Omega$ is toroidal in the 3-dimensional domain. $\Omega$ cannot be determined a priori. Its boundary can be defined by the smallest non-zero root of $r$ in equation $\xi(r, x_0)= 0$ at a given longitudinal position $x_0$. 

Here, we only consider the case of $A>0$, because $A<0$ gives rise to a monotonic increase of the integral in the $r$ direction and thus an infinite boundary extent. In this case, boundary condition (\ref{eq:BCa}$a$) requires that the axis of symmetry is a streamline, and without loss of generality, $B/A$ is chosen to be the streamfunction value. Because $W_{-\frac{\lambda^2}{4\alpha},\frac{1}{2}}(0)\neq 0$, we have to set $C_{\lambda 2}=0$ to prevent the stream function from varying at $r=0$. 

The boundary condition (\ref{eq:BCb}$b$) comes from the observation that the vorticity distribution typically peaks at the center, and the vortex shape is nearly symmetrical with respect to the plane of $x=0$. Note also that the real part of the Whittaker $M$ function $M_{-\frac{\lambda^2}{4\alpha},\frac{1}{2}}(\alpha r^2)$ is not integrable, and thus $C_{\lambda 1}$ has to be a pure imaginary number. Hence, the vortex ring solution can then be expressed as
\begin{equation}\label{eq:vtps}
\psi=\int_{0}^{\infty}iC(\lambda)M_{\frac{\lambda^2i}{4\sqrt{A}},\frac{1}{2}}(\sqrt{A}r^2i)\cos(\lambda x)d\lambda+B/A
\end{equation}
\begin{equation}\label{eq:vtxi}
\xi=A r\int_{0}^{\infty}iC(\lambda)M_{\frac{\lambda^2i}{4\sqrt{A}},\frac{1}{2}}(\sqrt{A}r^2i)\cos(\lambda x)d\lambda
\end{equation}
where $C(\lambda)$ is a real function that can be viewed as the modulus function. Because the integral is symmetric with respect to $\lambda=0$, the integral limits are constrained to the positive range. 

\begin{figure}
\minipage{0.4\textwidth}  
  \includegraphics[height=\linewidth]{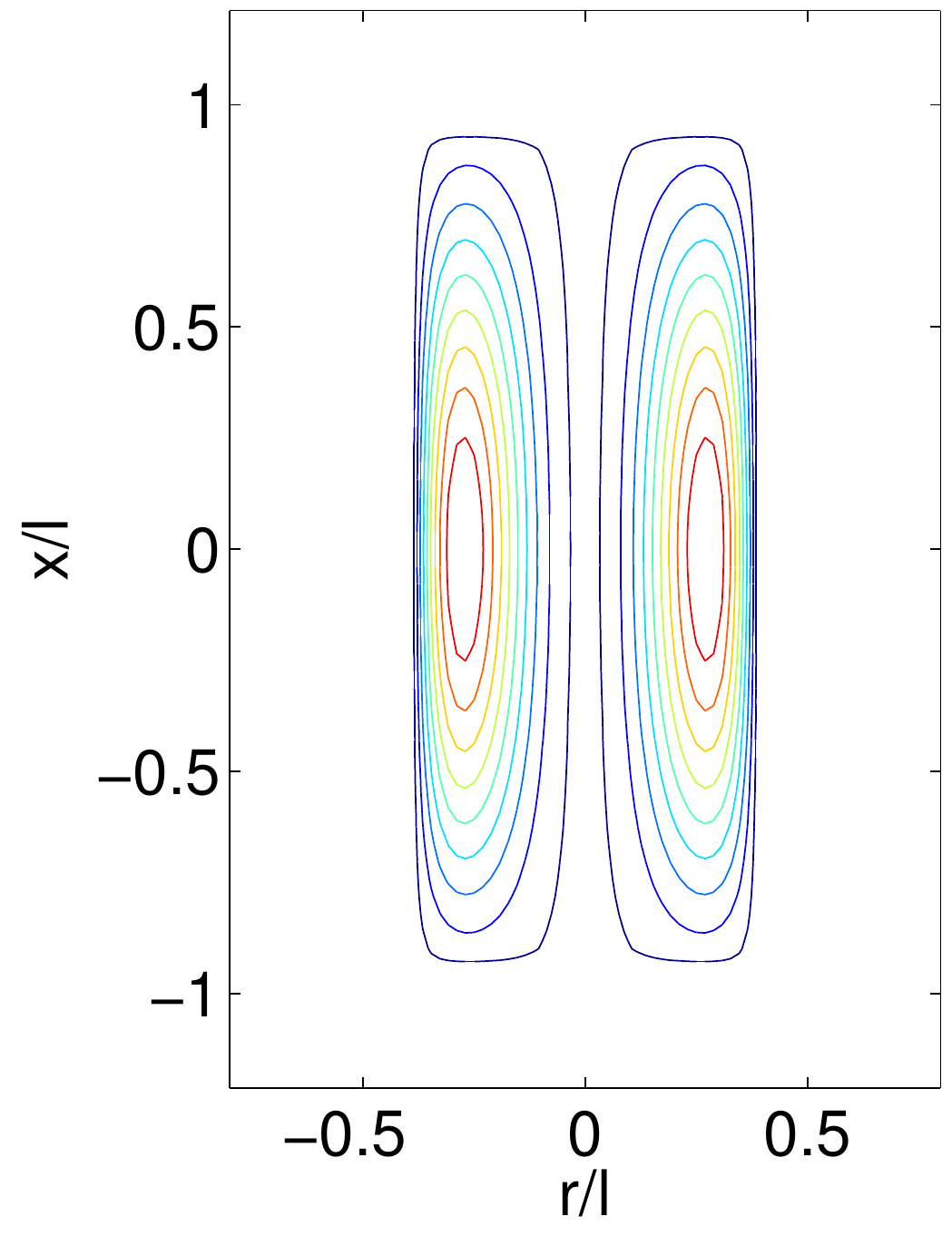}
  \caption{Streamlines for $\sigma=0.2$ and $\beta=0.39$}
  \label{fig:psi0p2}
\endminipage
\hfill
\minipage{0.58\textwidth}  
  \includegraphics[height=0.7\linewidth]{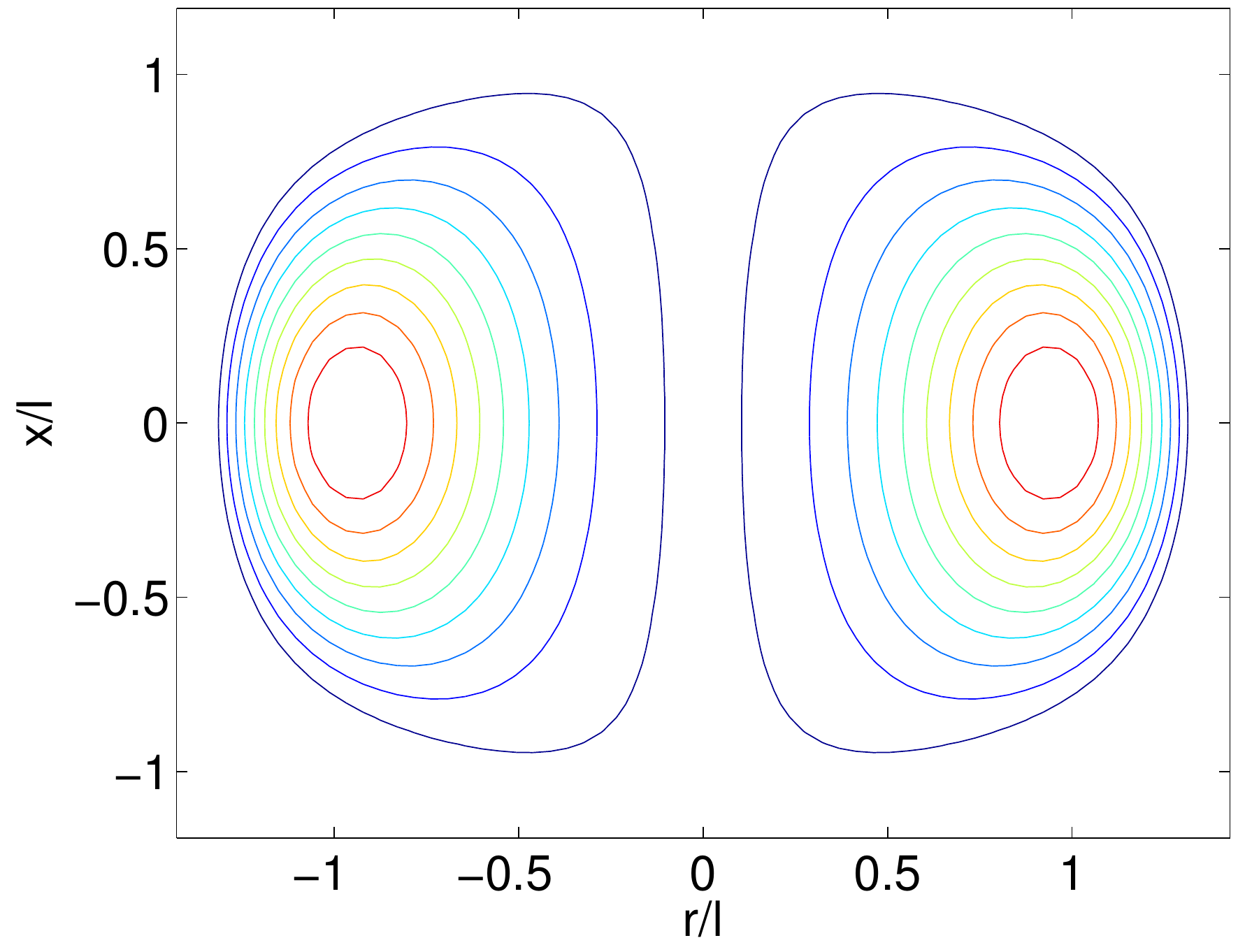}
  \caption{Streamlines for $\sigma=0.6$ and $\beta=1.32$}
  \label{fig:psi0p6}
\endminipage
\end{figure}

$C(\lambda)$ needs to be resolved in order to obtain the solution. Mathematically, if $C(0)\neq0$, the equation is reduced to a parallel shearing flow, which has no finite boundary in the $x$ direction; and if $C(\infty)\neq0$, the integral is not finite. Therefore, for a closed boundary and for integrability, $C(\lambda)$ has to vanish as $\lambda\to0$ or $\infty$. Physically, $C(\lambda)$ can be seen as a spectrum of vortex length scales, which has reached a self-balanced state after the vortex ring formation. Hence, it is reasonable to assume that $C(\lambda)$ is continuous and smooth. Although many functions are able to satisfy the conditions mentioned above, the simplest candidate would be the Rayleigh distribution function, i.e.
\begin{equation}\label{eq:rayl}
C(\lambda)=\frac{\lambda }{{{\sigma ^2}}}{e^{ - \frac{{{\lambda ^2}}}{{2{\sigma ^2}}}}},
\end{equation}
where $\sigma$ is a constant. Taking $A=1$ and $B=0$, we reach a normalized form of equation (\ref{eq:vtps}), i.e.
\begin{equation}\label{eq:standard1}
\psi=\int_{0}^{\infty}i\frac{\lambda }{{{\sigma ^2}}}{e^{ - \frac{{{\lambda ^2}}}{{2{\sigma ^2}}}}}
M_{\frac{\lambda^2}{4}i,\frac{1}{2}}(l^2 r^2i)\cos(\lambda l x)d\lambda.
\end{equation}
Accordingly, the vorticity distribution becomes
\begin{equation}\label{eq:standard2}
\xi=r\int_{0}^{\infty}i\frac{\lambda }{{{\sigma ^2}}}{e^{ - \frac{{{\lambda ^2}}}{{2{\sigma ^2}}}}}
M_{\frac{\lambda^2}{4}i,\frac{1}{2}}(l^2 r^2i)\cos(\lambda l x)d\lambda.
\end{equation}
%
For illustration, two typical vortex ring patterns are shown in figures \ref{fig:psi0p2} and \ref{fig:psi0p6}. Note that the pattern has been further normalized by the length scale $l$ so that the $x$ extent of the vortex ring is within the range from -1 to 1. Comparing figures \ref{fig:psi0p2} and \ref{fig:psi0p6}, the aspect ratio, $\beta=r_{max}/z_{max}$, is greater with a greater $\sigma$, and the boundary is more rectangular.

Generally, equations (\ref{eq:standard1}) and (\ref{eq:standard2}) depict a family of new vortex ring models. In the following, we obtain the properties of the present model by numerical methods.

\section{Properties of the vortex ring family}\label{sec:properties}

The current model of equation (\ref{eq:standard1}) has only one free parameter, $\sigma$, by which a family of vortex rings can be modeled. The basic properties of this model can be described by a series of integrals of the vortex ring dynamics including the circulation ($\Gamma$), the hydrodynamic impulse ($I$), and the energy ($E$), which are defined as below:
\begin{equation}\label{eq:GIE}
\Gamma  = \int_0^\infty  {\int_{ - \infty }^\infty  {\xi dxdr} }, \qquad
I = \pi \int_0^\infty  {\int_{ - \infty }^\infty  {{r^2}\xi dxdr} }, \qquad
E = \pi \int_0^\infty  {\int_{ - \infty }^\infty  {\xi \psi dxdr} } .
\end{equation}

\begin{figure}
\minipage{0.48\textwidth}  
  \includegraphics[width=\linewidth]{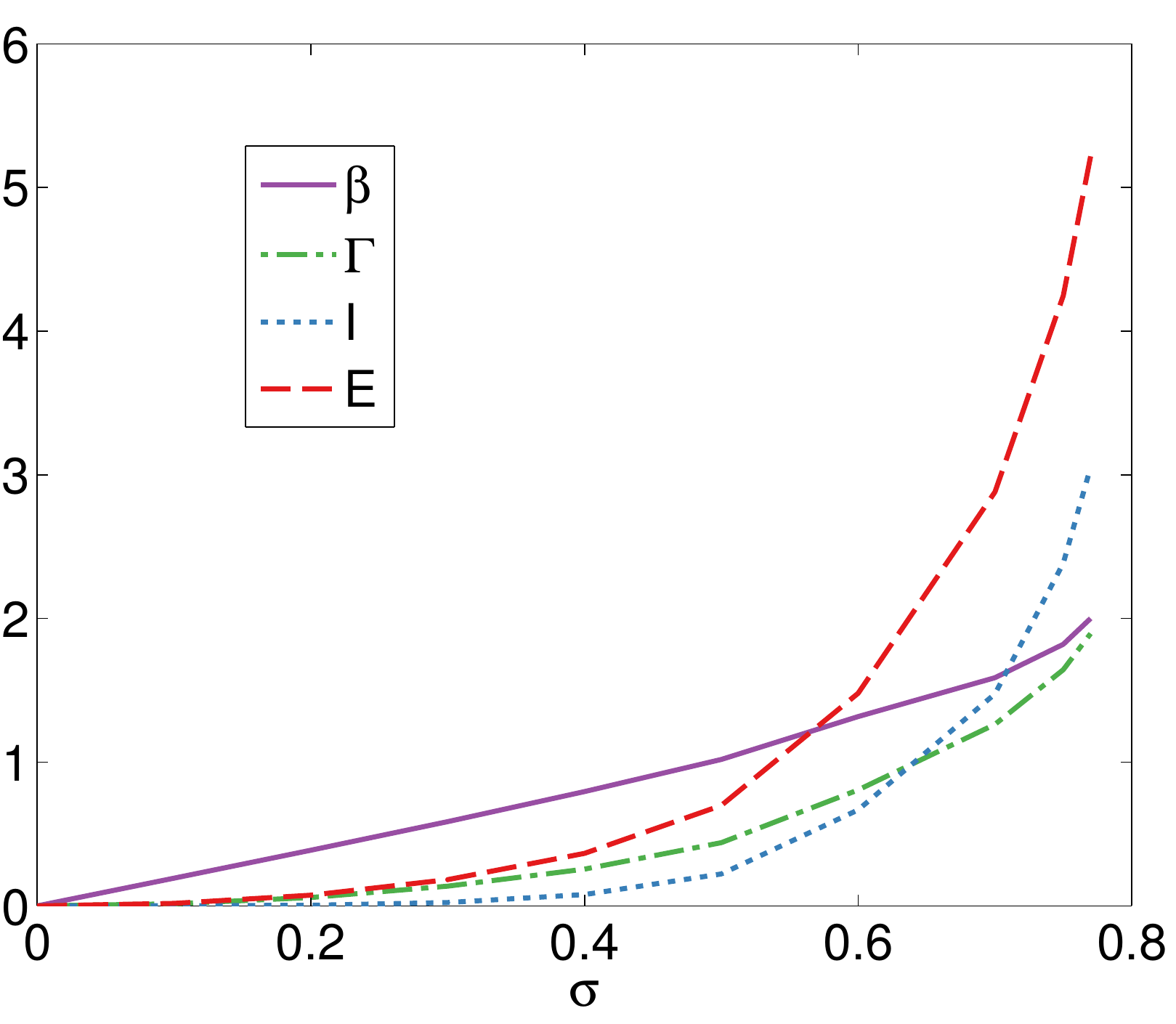}
  \caption{The aspect ratio, $\beta$, the circulation, $\Gamma$, the impulse $I$ and the energy $E$ against the free parameter, $\sigma$}
  \label{fig:beta_sigma}
\endminipage
\hfill
\minipage{0.48\textwidth}  
  \includegraphics[width=\linewidth]{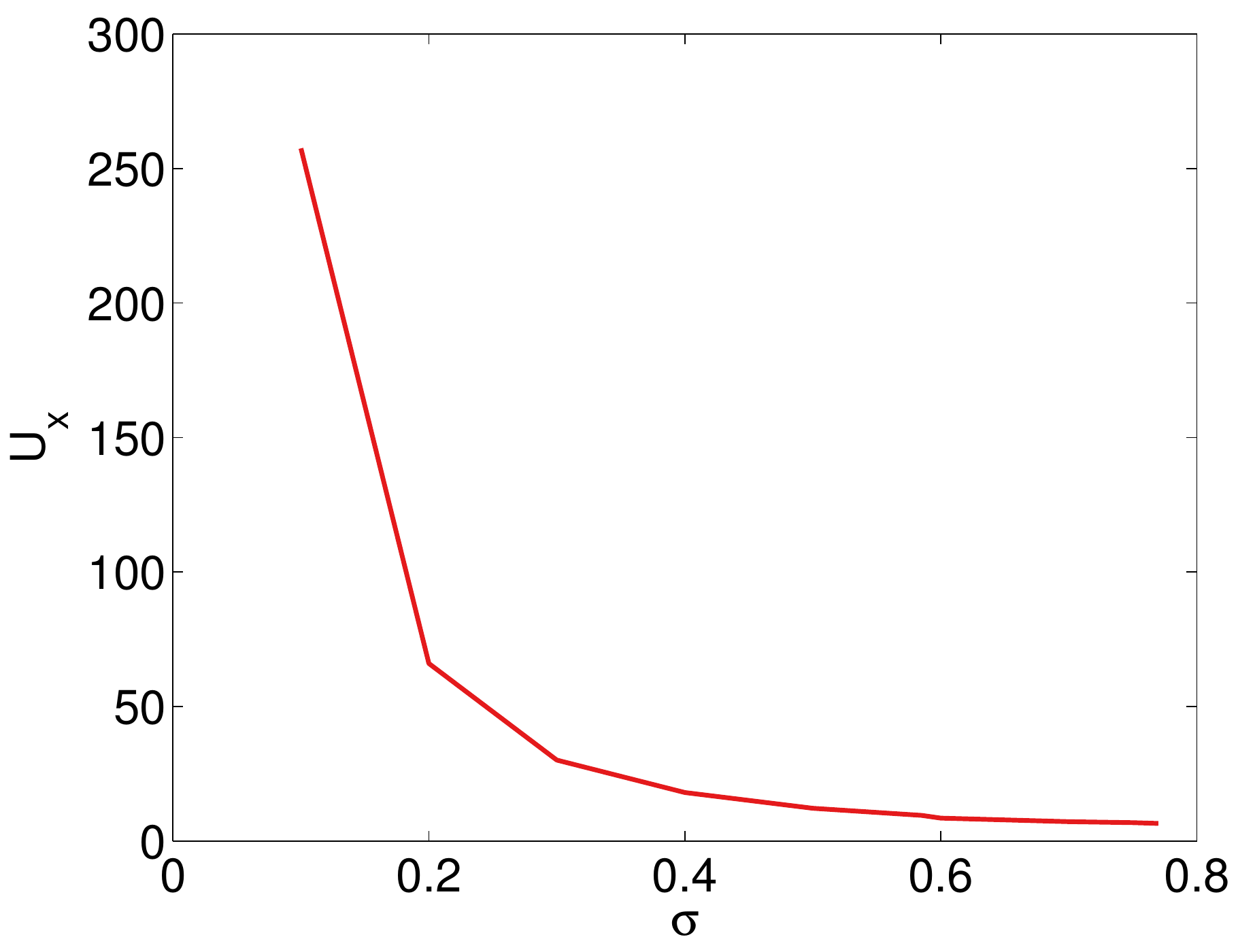}
  \caption{The variation of the propagation velocity, $U_x$, against the free parameter, $\sigma$}
  \label{fig:Ux}
\endminipage
\end{figure}

By varying $\sigma$, we can derive the variation of $\beta$, $\Gamma$, $I$, and $E$, which are shown in figure \ref{fig:beta_sigma}. The aspect ratio and all the integrals generally increase with larger $\sigma$. Before $\sigma=0.7$, the aspect ratio $\beta$ is increasing almost linearly. Afterwards it increases dramatically and approaches infinity at $\sigma=\sim0.78$. The integrals increase much more quickly than $\beta$ and also approaches infinity at the same place. Note that when $\sigma=0$, the $r$ extent approaches zero (or $l\to\infty$), and thus the vortex ring is reduced to an infinite line. As a result, all the integrals also vanish.

A useful parameter in vortex ring dynamics is the propagation velocity, $U_x$, which is defined earlier in this paper. Theoretically, it can be estimated by the following general equation \citep{helmholtzintegrals, lamb1993hydrodynamics}, 
\begin{equation}
{U_x} = \frac{{\int_0^\infty  {\int_{ - \infty }^\infty  {\left( {\psi  - 6x\frac{{\partial \psi }}{{\partial x}}} \right)\xi dxdr} } }}{{\int_0^\infty  {\int_{ - \infty }^\infty  {{r^2}\xi dxdr} } }}.
\end{equation}
It is difficult to derive an explicit expression for $U_x$. Hence, we use a numerical method to compute $U_x$, which is shown in figure \ref{fig:Ux}. From this figure, the propagation velocity $U_x$ decreases with larger $\sigma$, but the slope of the decrease is much reduced when $\sigma$ is near $\sim0.78$. At $\sigma\to0$, $U_x\to\infty$. Note that the results in figure \ref{fig:Ux} are normalized. The absolute values depend on the scaling.

\begin{figure}
\minipage{0.48\textwidth}  
  \includegraphics[width=\linewidth]{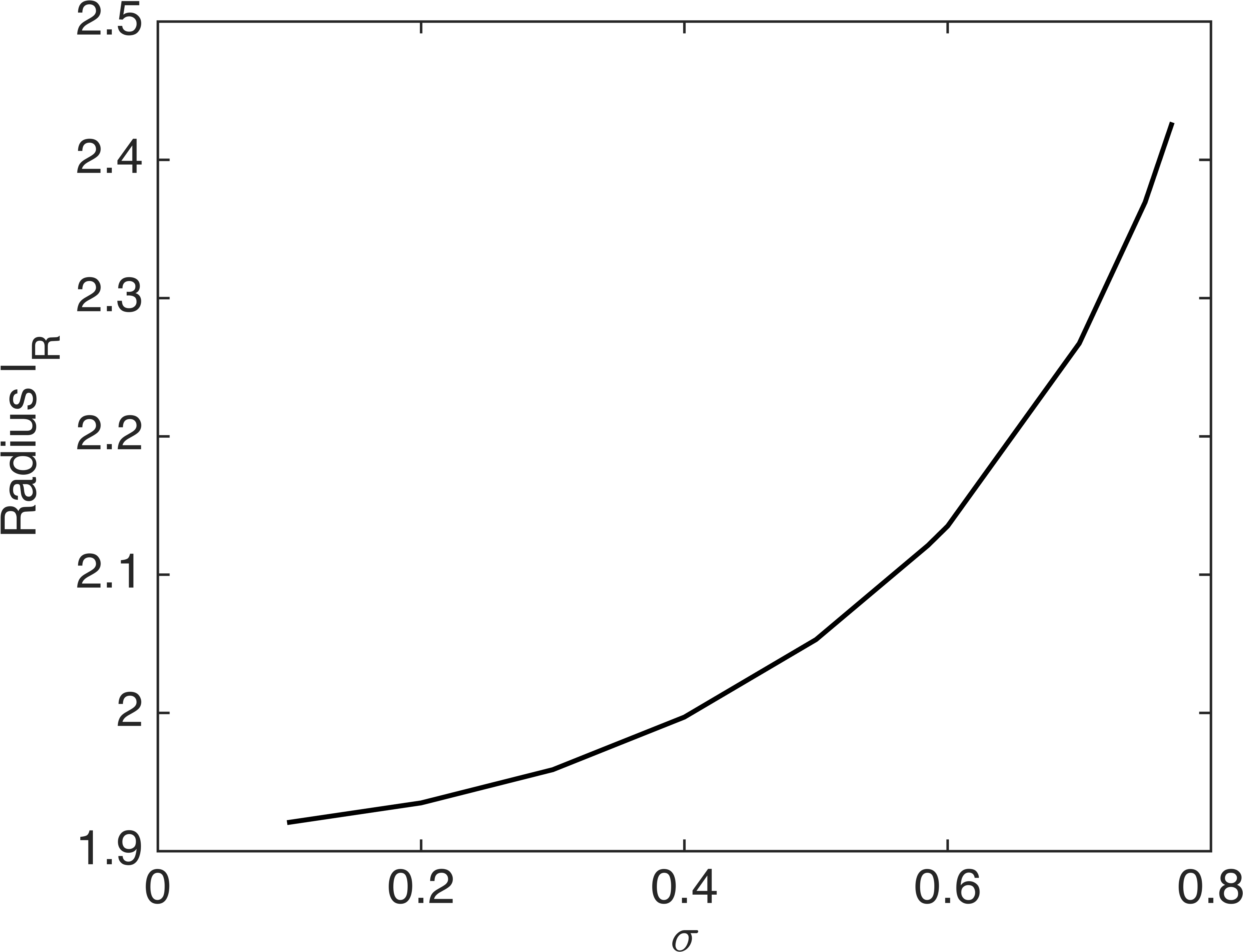}
  \caption{The variation of vortex ring radius, $l_R$, against the free parameter, $\sigma$}
  \label{fig:sigma_R}
\endminipage
\hfill
\minipage{0.48\textwidth}  
  \includegraphics[width=\linewidth]{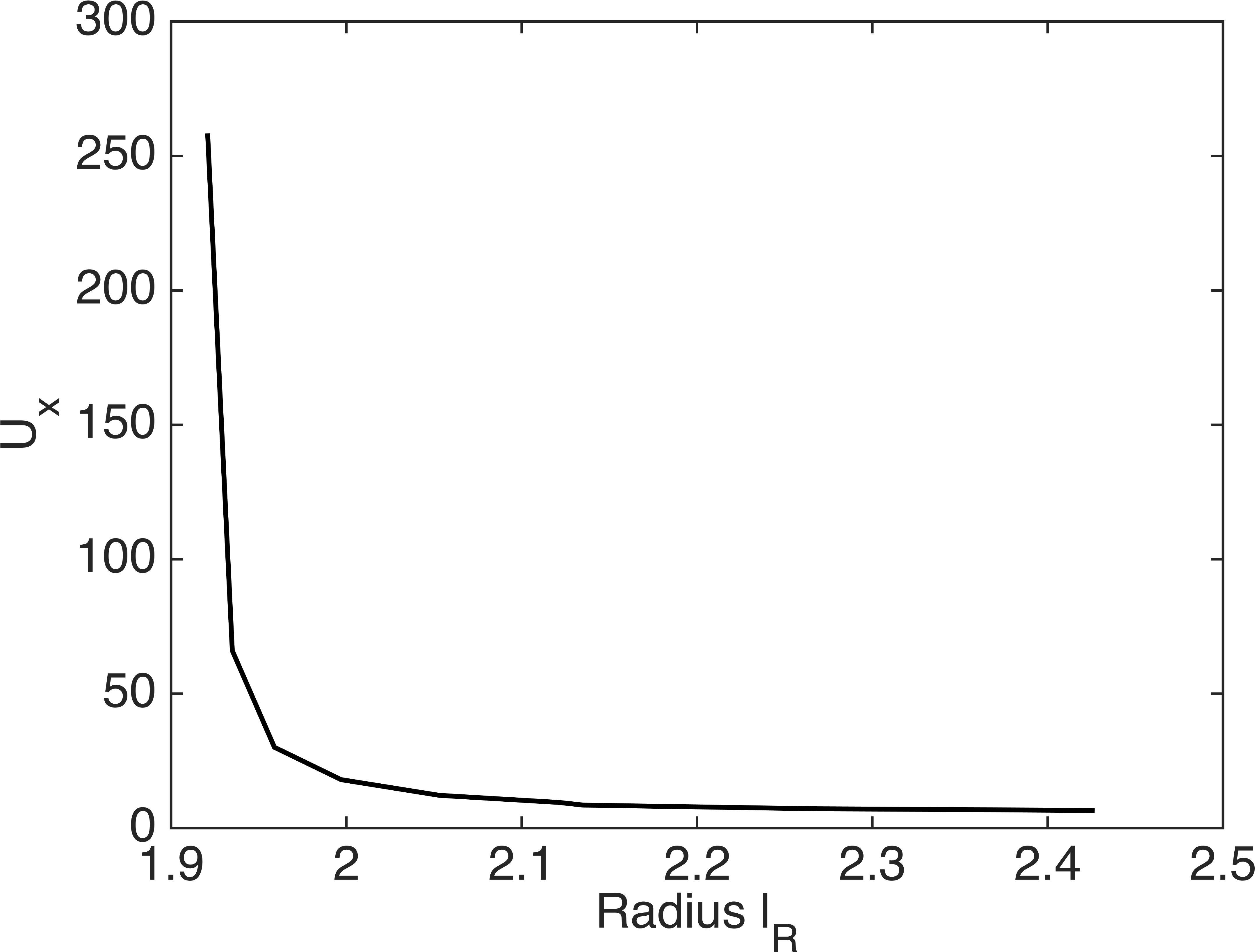}
  \caption{The variation of the propagation velocity, $U_x$, against the radius, $l_R$}
  \label{fig:R_Ux}
\endminipage
\end{figure}

The normalized radius, $l_R$ (defined later in section 5), can be derived as a function of $\sigma$ (figure \ref{fig:sigma_R}) and its relationship to the translation velocity $U_x$ can be found in figure \ref{fig:R_Ux}. We can find that the radius increases with $\sigma$ exponentially and the translation velocity decrease with the radius quickly at small radius and slowly later.

Also note that the effect of $A\neq1$ can be incorporated by a correction to the aspect ratio $\beta$ and the length scale $l$. If $B\neq0$, the aspect ratio of the streamlines remains, but the vorticity  and other integrals will be increased accordingly.

\section{Comparison to numerical simulations}\label{sec:Comparison_numerical_simulations}
The present solution takes advantages that at high $\Rey$ laminar flow, the viscous effects are present but negligible. This enables us to make a meaningful comparison with corresponding direct numerical simulations in this range.

To examine the properties of vortex rings, \cite{Danaila2008} performed axisymmetric Direct Numerical Simulations of a piston generated vortex ring and compared the results with the Norbury-Fraenkel family and Kaplanski and Rudi's analytical solutions. Their simulations are based on the formation number of $L/D=4$ which is close to that of a saturated vortex ring. The comparison included the streamlines and vorticity distribution at $tU_0/D=30$, which are shown in figure \ref{fig:k5}$(a)-(f)$. 

\begin{figure}
  \centerline{\includegraphics[width=0.8\linewidth]{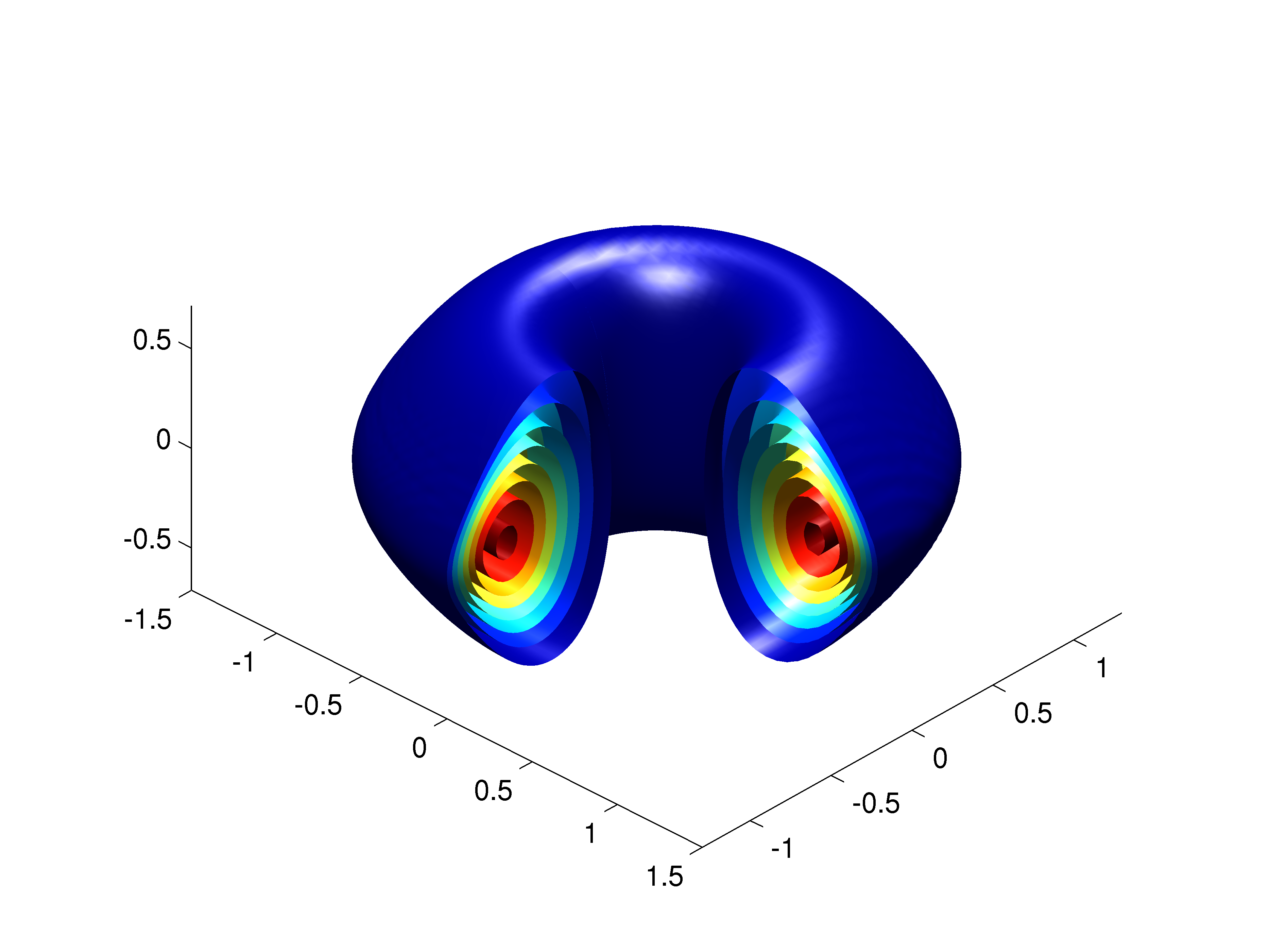}}
  \caption{Three-dimensional vortex ring structure of vorticity.}
\label{fig:3d}
\end{figure}

\begin{figure}
  \centerline{\includegraphics[width=\linewidth]{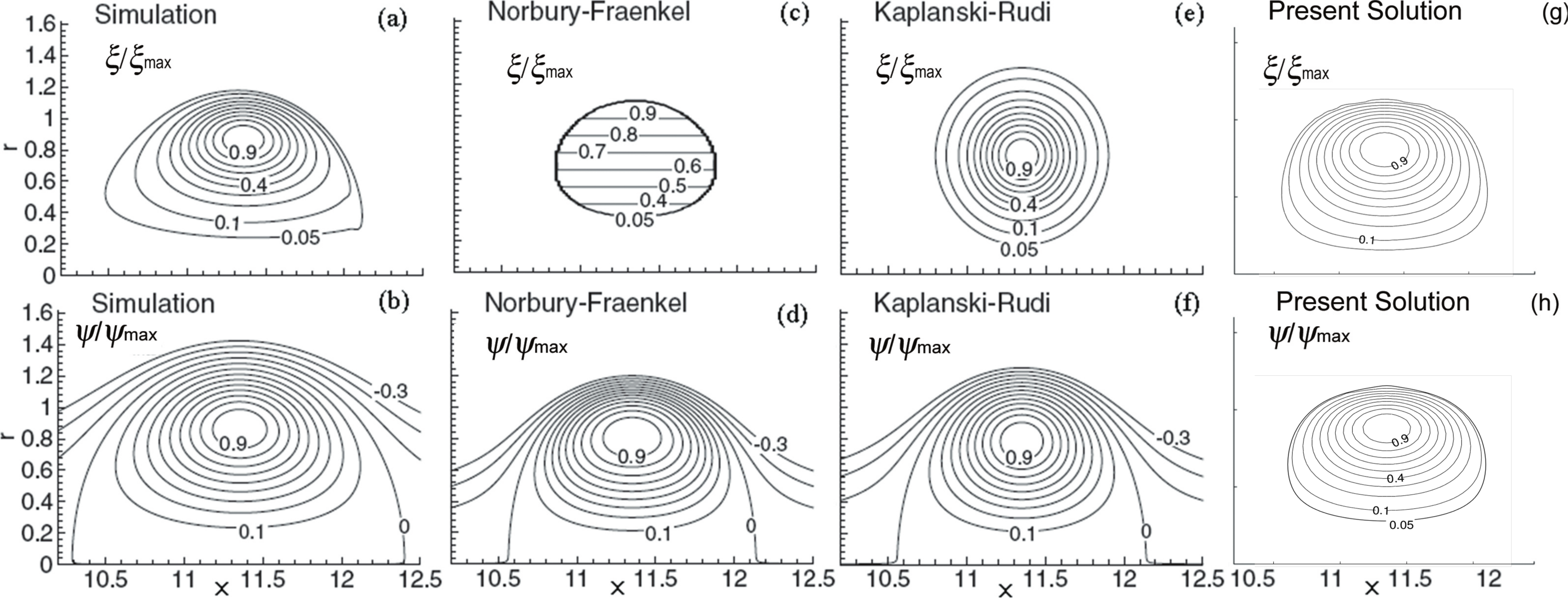}}
  \caption{Comparison of the ideal models of Norbury-Fraenkel, Kaplanski-Rudi, and the present solution with the direct numerical simulations by\cite{Danaila2008}. The upper row is the vorticity distribution and the lower is the streamline. (a)-(f) modified from \cite{Danaila2008}.}
\label{fig:k5}
\end{figure}

To quantify the characteristics of the vortex ring using the present approach, the parameter, $\sigma$, has to be specified. By close examinination, the aspect ratio of DH is $\beta=1.23$, which corresponds to $\sigma=0.585$ in the present model from figure \ref{fig:beta_sigma}. Furthermore, the normalized standard form should be adjusted in its magnitude to compare with the simulation results, with $C(\lambda)$ scaled by a factor of 0.65 and $l=0.94$. Using this configuration, we can visualize the three dimensional structure of the vortex ring in figure \ref{fig:3d}. The vortex ring has a torus shape with a deformed cross-section. The outside of the vortex ring is narrower than the inner side. The toroid is also sliced to show the cross-section, where the vorticity distribution can be seen. In the following, we draw a series of comparisons to show that  the present results agree well with the numerical simulations. 

Overall, figure \ref{fig:k5} shows that the present solution provides a better agreement than the other peer models. A close match of streamlines and vorticity distributions is obtained as shown in figure \ref{fig:k5} (g) and (h). The streamlines of figure \ref{fig:k5} (b), (d), (f) and (h) illustrate that all models essentially share similar characteristics, but the present solution in (h) achieves better matching. More notably, the vorticity distribution comparison demonstrates a significant improvement.  As discussed before, the Norbury-Fraenkel's family has a major shortcoming due to the unrealistic linear vorticity distribution. The Kaplanski-Rodi's model improves with a continuous vorticity distribution at the boundaries, but the vorticity contours dictate a circular shape which is also unrealistic. The present solution better depicts the asymmetry of the vorticity distribution by accommodating multiple harmonics. The small discrepancy at the boundary might be attributed to the absence of viscous effects.

\begin{figure}
\minipage{0.5\textwidth}
  \includegraphics[width=\linewidth]{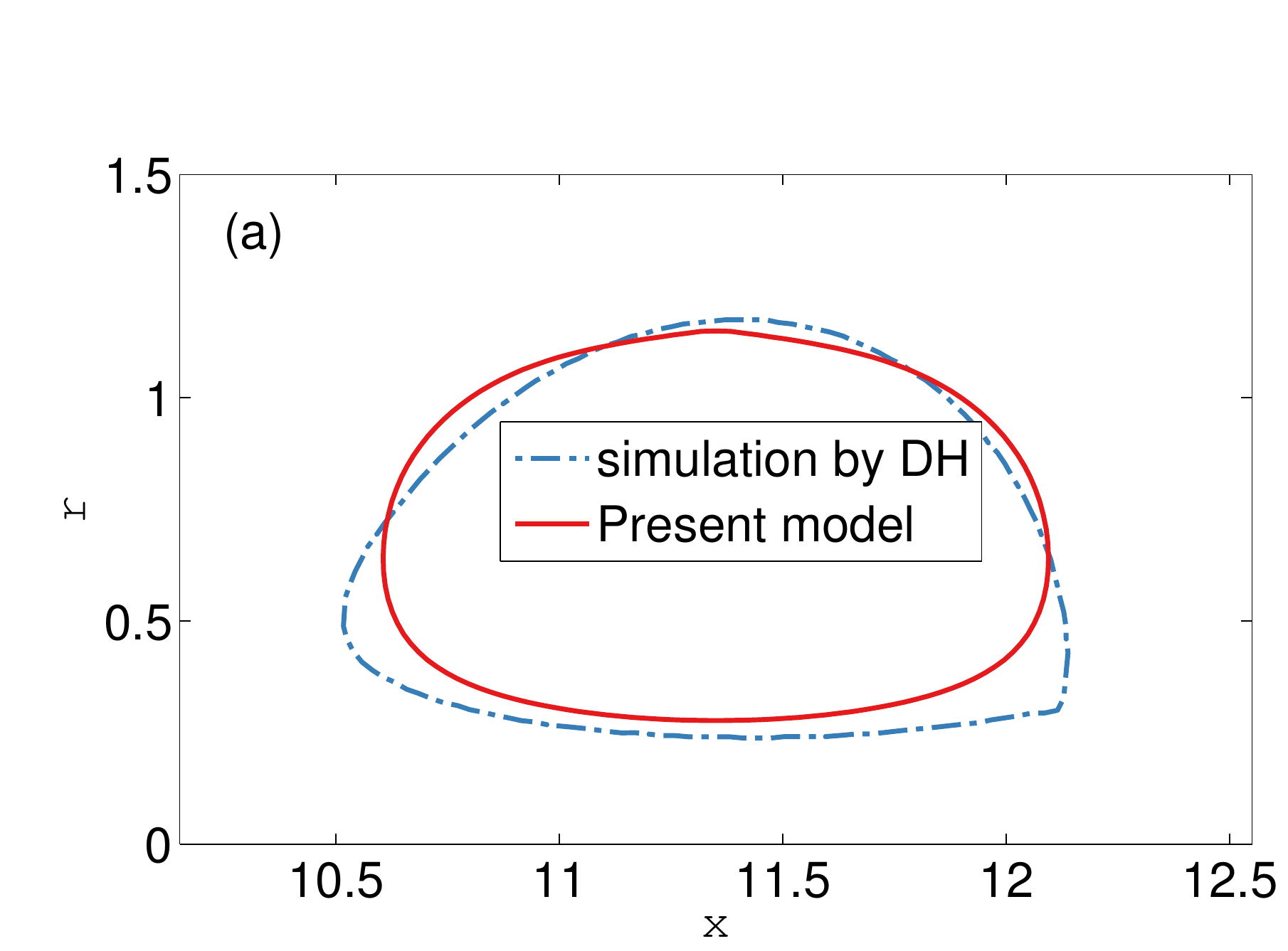}
\endminipage
\hfill
\minipage{0.5\textwidth}
  \includegraphics[width=\linewidth]{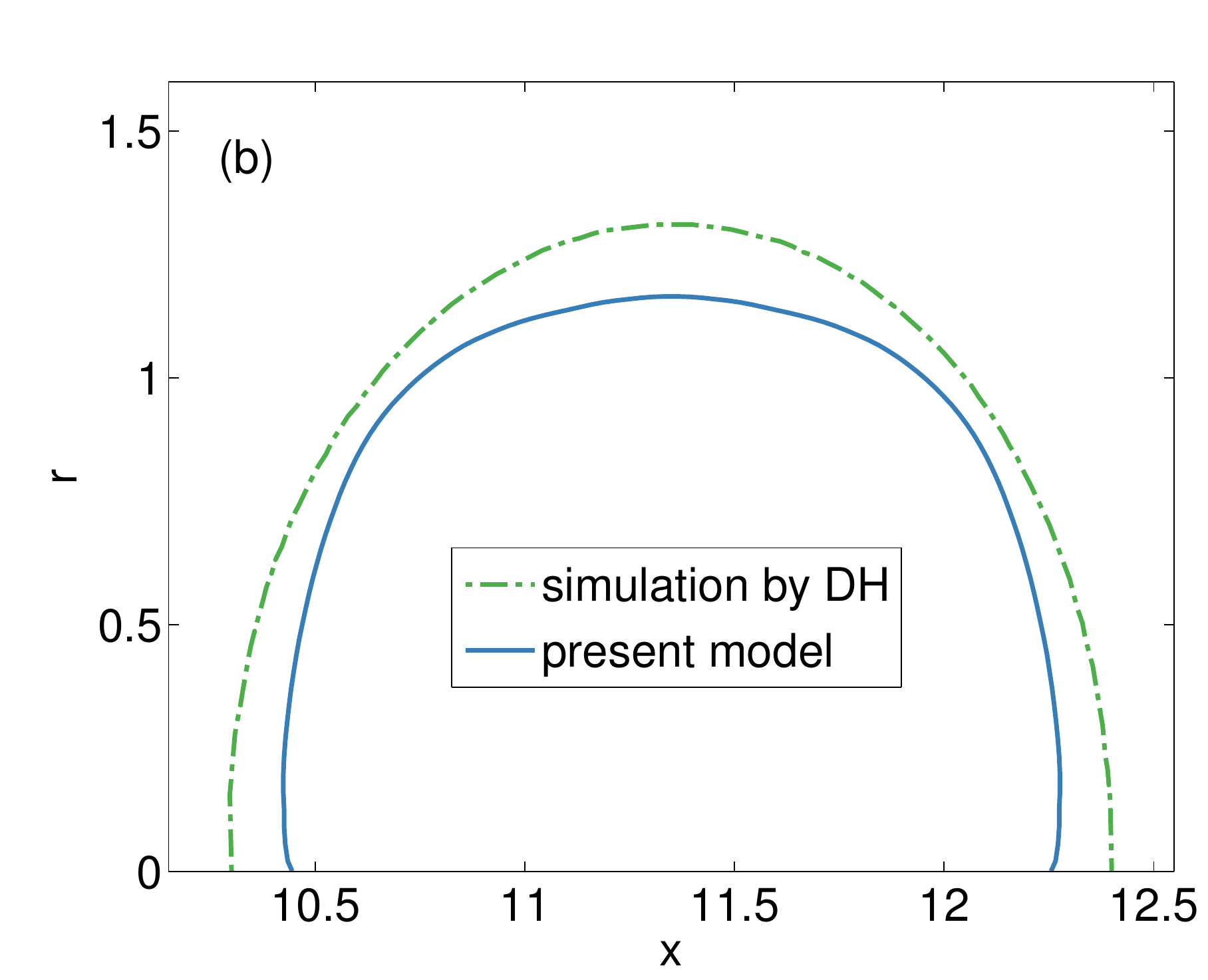}
\endminipage
  \caption{Comparison of boundary contours with DH: (a) the vortex core and (b) vortex bubble. }
\label{fig:k6}
\end{figure}

With the present solution, the boundaries of the vortex cores and vortex bubbles are also found to better match the numerical simulations by \cite{Danaila2008}. The isocontour of $5\%$ of the maximum vorticity is taken as the vortex core boundary shown in figure \ref{fig:k6}(a). The present solution shares similar characteristics of a sharper top and flatter bottom with the numerical results. In addition, the outline of the vortex bubble where the stream function decays to zero is shown in figure \ref{fig:k6}(b). With the appropriate $\sigma$, the present solution has the same aspect ratio as DH, but underestimates slightly the $r$ and the $x$ extents, which may again be due to the fact that the viscous effect is neglected. Although the cross-sectional area is underestimated by $18\%$, this is a considerable improvement over the peer models in which $\sim40\%$ or more underestimation is common.

\begin{figure}
\minipage{0.45\textwidth}  
  \includegraphics[width=1.2\linewidth]{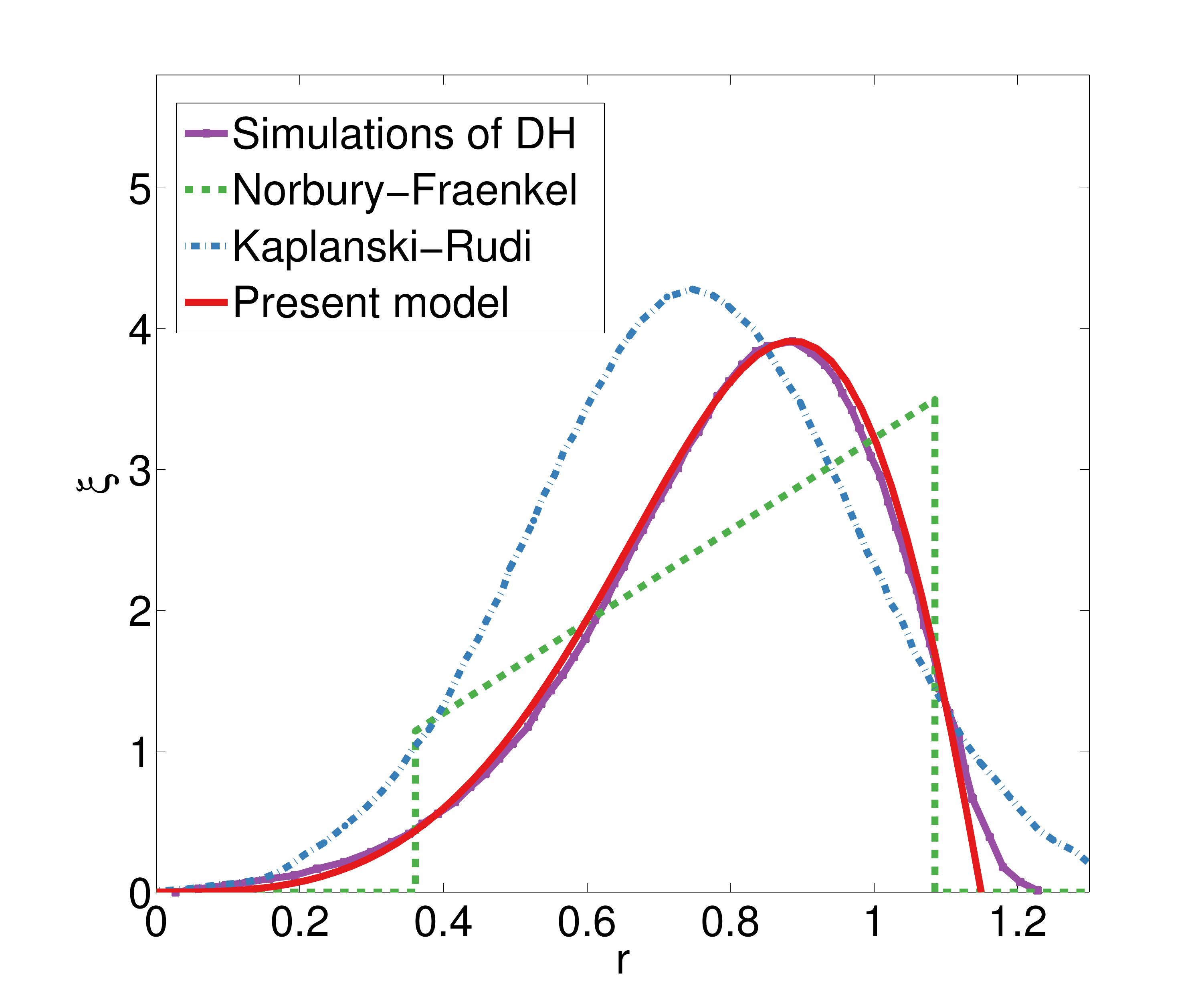}
  \caption{Radial distribution of vorticity through the vortex center: comparison to different vortex ring models }
  \label{fig:k7}
\endminipage
\hfill
\minipage{0.45\textwidth}  
  \includegraphics[height=5cm,width=1\linewidth]{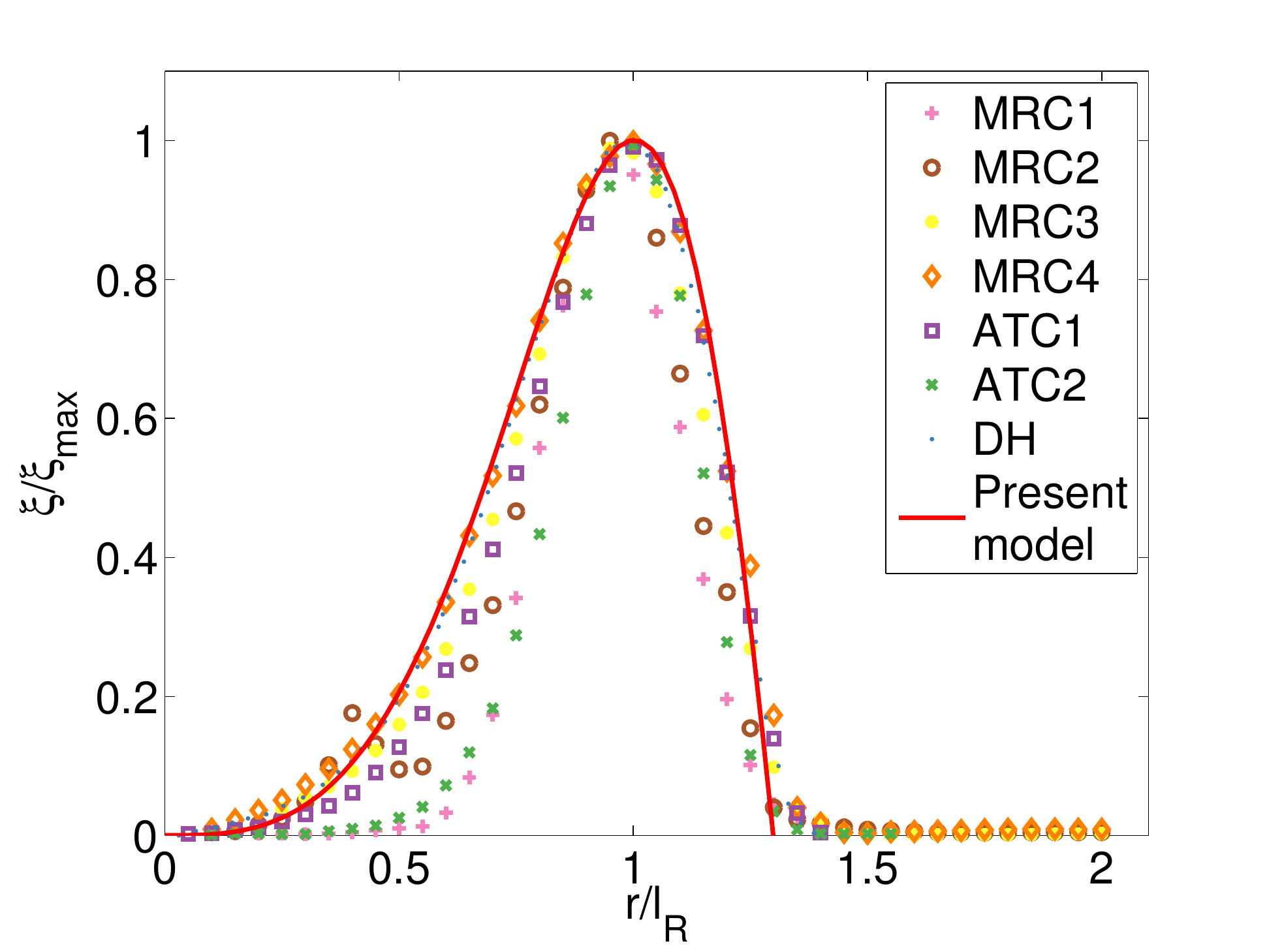}
  \caption{Radial distribution of vorticity through the vortex center: comparison to different numerical simulations in literature. Legend entries can be found in table \ref{tab:kd}. }
  \label{fig:k8}
\endminipage
\end{figure}

The radial distribution of vorticity through the vortex center is shown in figure \ref{fig:k7}. The numerical results show a smooth peak profile, with a long tail expanding to the axis of symmetry and a steep slope at the outside. Quantitative comparison reveals that the Norbury-Fraenkel family has an unrealistic linear profile; while the Kaplanski-Rudi model features a symmetrical Gaussian distribution which by definition cannot reproduce the asymmetry observed in the numerical results.

Besides the good comparison with \cite{Danaila2008}, similar qualitative agreement in vorticity distribution contours is also obtained with observations reported in the literature, e.g.\ \cite{Zhao2000}. However, for quantitative comparison, to the authors' knowledge, only \cite{Mohseni2001} (MRC) and  \cite{Archer2008} (ATC) offered high quality data that can be compared alongside \cite{Danaila2008} (DH). Both MRC and ATC also adopted Direct Numerical Simulations to reproduce the vortex ring flows similar to DH. Specifically, MRC focused on the pinch-off process. The radial vorticity distributions through the vortex ring center at different times are shown as MRC1$\sim$MRC4 in figure \ref{fig:k8}. In contrast, ATC simulated the transition of vortex rings from laminar to turbulent, and only a snap shot of the radial vorticity distribution in the laminar range for both thin and thick cores were reported. They are shown as ATC1 and ATC2 respectively in figure \ref{fig:k8}. 

The maximum value, $\xi_{max}$, and the peak radius, $l_R$, can be extracted directly from MRC and ATC. They are used as the vorticity and length scales for normalization of their reported data. Meanwhile, we show the present solution in figure \ref{fig:k7} using the best fitting coefficient $\sigma=0.585$ with DH, but with the output of the results normalized by $\xi_{max}$ and $l_R$. In figure \ref{fig:k8}, clearly the present solution compares well with the numerical results by MRC and ATC (the abbreviations and their corresponding flow conditions can be found in table \ref{tab:kd}). Specifically, MRC reported their vorticity distributions as a time series of a specific simulation. In their case, the distribution was narrower in earlier time when the vortex was still developing, and became wider later on when $U_x$ approached constant. The latter is closer to our steady state assumption in the derivation of (\ref{eq:sxps}) (see the constant slope of the vortex center position in figure 1 of MRC). Strictly speaking the present solution is only appropriate to be compared with the steady state MRC4, where an excellent comparison is found. Note that in MRC2, a secondary peak can be identified near the axis of symmetry, reflecting the merging trailing stem.  

ATC observed in the late laminar phase before turbulent onset that ``the vorticity profile across the core region relaxes towards a new equilibrium state, when the axisymmetric inviscid ideal is solely a function of the stream function $\psi$; the distribution becomes skewed, decreasing faster toward the bubble edge than the ring centre.'' This observation is consistent with the present model development, and validates our comparison exercise. Of the two cases in ATC, one was with a relatively saturated state of development shown as ATC1, and the other a thin core vortex ring with a much narrower peak that had not yet reached the saturated phase shown as ATC2. Therefore, it's more appropriate to compare with ATC1 instead of ATC2. As expected, the present model, which is derived with the steady state assumption, matches well the former. Finally, DH is also normalized and shown in figure \ref{fig:k8} for comparison. To summarize, the present solution is able to match closely the DNS results of laminar vortex rings with high $\Rey$ reported in the literature when the post-formation stage is approached, which covers a wide range of $\Rey=1400-5500$.

\begin{figure}
  \centerline{\includegraphics[width=0.8\linewidth]{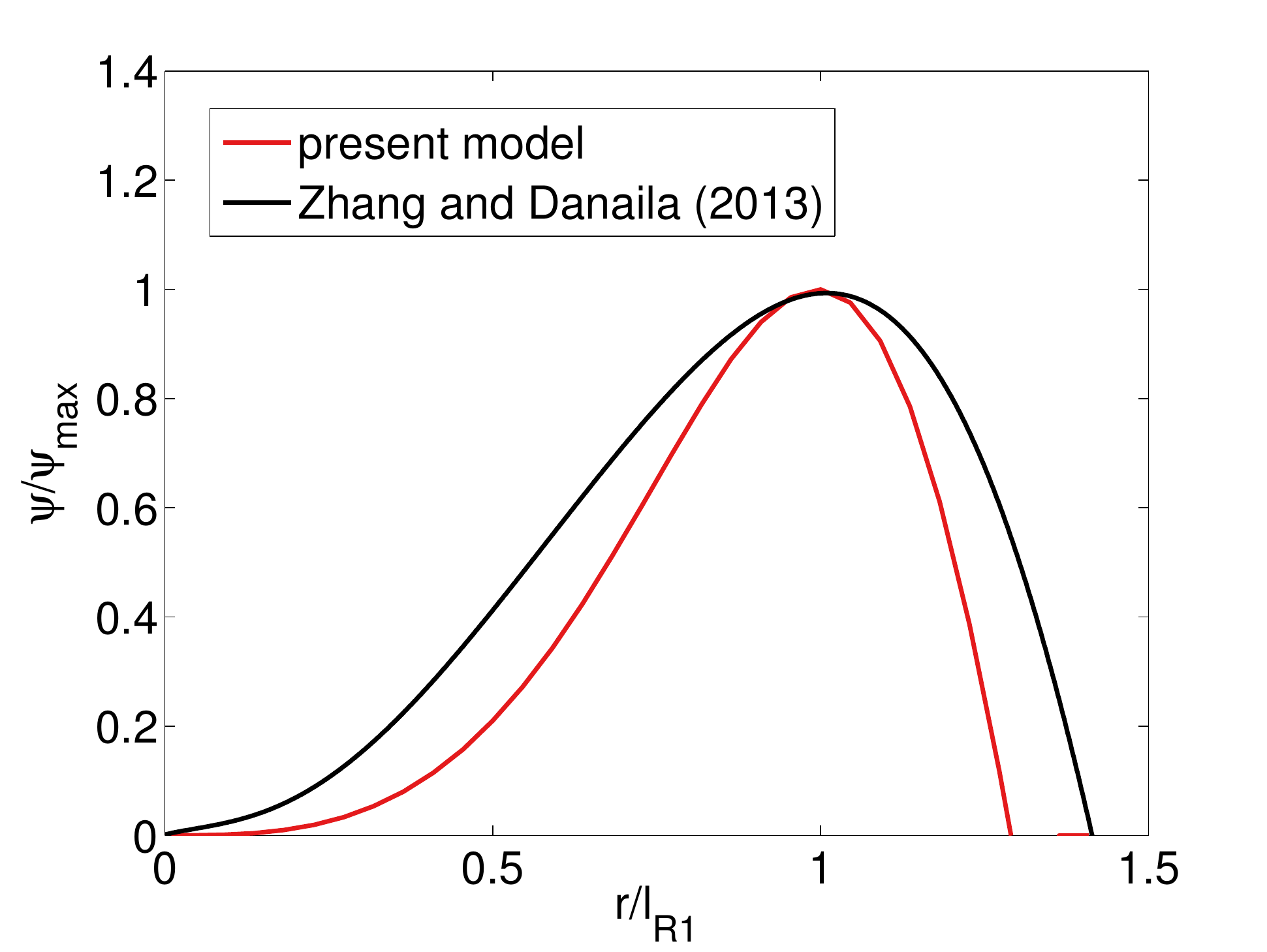}}
  \caption{Radial distribution of the stream function through the vortex center compared to \cite{zhang2013existence}.}
\label{fig:k9}
\end{figure}

Recently, \cite{zhang2013existence} derived a family of vortex rings with varying levels of saturation. By examining the elliptical boundaries, they proved the existence of the elliptical vortex rings. Using numerical methods, they offered data on the stream function distribution in the radial direction. Corresponding to their most saturated situation, a comparison is drawn in figure \ref{fig:k9} after normalization with the radius and the value of the maximum of $\psi$, i.e. $l_{R1}$ and $\psi_{max}$. Although the present solution is relatively narrower, the shapes of these profiles agree well with the results from \cite{zhang2013existence}. The slight discrepancy may be due to the fact that there is no superimposed potential flow in the present solution.

Although the current paper is targeted to model a high Reynolds number vortex ring, it can also be applied to truly inviscid vortex rings, which are currently investigated theoretically and experimentally in superfluid and atomic Bose-Einstein condensates, as reviewed for example by \cite{barenghi2009vortex}.

\section{Summary and conclusions}\label{sec:summary_conclusion}
The present study proposes a theoretical model of a family of steady state vortex rings. First, a solution to the Grad-Shafranov equation from magnetohydrodynamics is derived, which is more succinct and compact compared with the existing solution by  \cite{Atanasiu2004}. Using the analogy between the equations of hydrodynamics and magnetohydrodynamics, a theoretical model to high $\Rey$ laminar vortex rings is developed based on this solution, which can satisfy the D-shaped boundary conditions that are commonly used in magnetohydrodynamics and suitable for vortex rings. A new family of theoretical vortex rings is derived based on this model with a free parameter $\sigma$. A numerical method is then used to calculate the properties of this family, including aspect ratio, circulation, impulse, energy, and propagation velocity. By choosing $\sigma=0.585$, a good match is obtained comparing the results from the present model and direct numerical simulations. The present model is found to be superior to the Norbury-Fraenkel family by virtue of its improved accuracy in the vorticity distribution and streamline pattern. Most importantly, the asymmetry and elliptical outline of the vorticity profile are well captured. Further comparison is performed with the normalized vorticity and stream function distributions reported in the literature and the present solution. Again, close agreement is obtained. The improved matching with the present solution can yield better assessment and models for engineering applications. \\

\section{Acknowledgement}

This research was supported by the National Research Foundation Singapore through the Singapore MIT Alliance for Research and Technology's CENSAM IRG research programme. The authors would like to thank the anonymous reviewers for comments to improve this manuscript.


\bibliographystyle{jfm}
\bibliography{AMM_submission}

\clearpage

\clearpage

\begin{table}
  \begin{center}
\def~{\hphantom{0}}
  \caption{Selected numerical results in the literature}

  \begin{tabular}{lccc}
  \hline
      Abbreviation  & Literature   			& \Rey	& $x/D$ \\[3pt]
      \hline
       MRC1         & Mohseni \etal (2001) 	& 3800 	& 3  \\
       MRC2   		& 						& 		& 5.8\\
       MRC3   		& 						& 		& 12.4\\
       MRC4   		& 						& 		& 19.9\\
       ATC1   		& Archer \etal (2008)	& 5500		& -\\
       ATC2   		& 						& 		& -\\
       DH         	& Danaila and H\'{e}lie (2008) 	& 1400 	& 11.35  \\
       \hline
  \end{tabular}
  \label{tab:kd}
  \end{center}
\end{table}

\end{document}